\documentclass[aps,pre,floats,twocolumn,showpacs,superscriptaddress]{revtex4}

\usepackage{graphicx}
\usepackage{graphics}
\usepackage{amsmath}
\usepackage{amsfonts}
\usepackage{amssymb}
\usepackage{epstopdf}
\usepackage{makeidx}
\usepackage{epsfig}
\usepackage{bm}
\usepackage{times}
\usepackage[colorlinks=true,citecolor=blue,linkcolor=blue,linktocpage=true,pagebackref=false]{hyperref}
\usepackage{color,soul} 
\usepackage{float}
\usepackage[english]{babel}

\newcommand{\be}{\begin{equation}}
\newcommand{\ee}{  \end{equation}}
\newcommand{\ba}{\begin{eqnarray}}
\newcommand{\ea}{  \end{eqnarray}}

\begin{document}


\title{Replicator population dynamics of group ($n$-agent) interactions. Broken symmetry, thresholds for metastability and macroscopic behavior.}

\author{Emmanuel Artiges}
\affiliation{\'Ecole Normale Sup\'erieure de Lyon,
Universit\'e Claude Bernard Lyon I, 69342 Lyon Cedex 07, France}
\author{Carlos Gracia-L\'azaro}

\affiliation{Institute for Biocomputation and Physics of Complex
Systems (BIFI), University of Zaragoza, Zaragoza 50009, Spain}
\affiliation{Departamento de F\'{\i}sica Te\'orica. University of
Zaragoza, Zaragoza 50009, Spain}

\author{Luis Mario Flor\'{\i}a}

\affiliation{Institute for Biocomputation and Physics of Complex
Systems (BIFI), University of Zaragoza, Zaragoza 50009, Spain}

\affiliation{Departamento de F\'{\i}sica de la Materia Condensada,
University of Zaragoza, Zaragoza 50009, Spain}

\author{Yamir Moreno}


\affiliation{Institute for Biocomputation and Physics of Complex
Systems (BIFI), University of Zaragoza, Zaragoza 50009, Spain}

\affiliation{Departamento de F\'{\i}sica Te\'orica. University of
Zaragoza, Zaragoza 50009, Spain}

\affiliation{ISI Foundation, Turin, Italy}

\date{\today}

\begin{abstract}
We analyze from basic physical considerations the Darwinian competition for reproduction (evolutionary dynamics) of strategists in a Public Goods Game, the archetype for $n$-agent (group) economical and biological interactions. In the proposed setup, the population is organized into groups, being the individual fitness linked to the group performance, while the evolutionary dynamics takes place globally. Taking advantage of (groups) permutation symmetry, the nonlinear analysis of the ``mesoscale'' Markov phase space for many competing groups is feasible to a large extent, regarding the expected typicality of evolutionary histories. These predictions are the basis for a sensible understanding of the numerical simulation results of the agent (microscopic) dynamics. Potential implications of these results on model-related issues as, e.g. group selection, the role of ``social norms'', or sustainability of common goods, are highlighted in concise terms, before the conclusion. 
\end{abstract}
\pacs{ 02.50.Le,89.65.-s,87.23.Ge} 
\maketitle

\section{Introduction}

Understanding how cooperative behaviour emerges in different contexts remains an outstanding question in modern evolutionary science(s) \cite{smith1982evolution,pennisi2005did}. The challenge posed by the observation of cooperation, when selfish behaviour provides higher fitness, has been studied in many different contexts, from Biology \cite{chuang2009simpson,gore2009snowdrift} to Economics \cite{kagel2016handbook} and Sociology \cite{kollock1998social}.  Several mechanisms have been proposed to explain this ``cooperation dilemma''. Among these, direct \cite{trivers1971evolution} and indirect (e.g., reputation) \cite{fowler2005human,nowak1998evolution} reciprocity rely on the idea that the cooperative behaviour may be favoured by the probability of future interactions; what has been termed network (spatial, lattice) reciprocity \cite{nowak1992evolutionary}, on the other hand is rooted on the (beneficial) assortative effects of the topology of individual connections. Other proposals, as kin selection (inclusive fitness), that refers to cooperation as favoring the reproductive success of an agent's relatives, even at a cost to the survival or reproduction of the individual \cite{hamilton1964genetical}, or group selection (competing groups of individuals), got entangled with the old debate in Evolution on multi-level (scale) selection, remaining open the question of the effective replicator: the individual, the group, the clade, the selfish gene, etc \cite{wynne1962animal,wilson1975theory,wilson1994reintroducing}. 
 
Although most of the theoretical \cite{rapoport1965prisoner,doebeli2005models} and experimental \cite{andreoni1993rational,cooper1996cooperation,gracia2012heterogeneous} studies on cooperation have focused on pairwise interactions, many biological \cite{chuang2009simpson}, social \cite{kagel2016handbook}, or economic \cite{kollock1998social} systems which are interesting from the perspective of cooperation involve $n$-agent (group) interactions, and perhaps the Public Goods Game (PGG) is one of the simplest, and most studied, ``group interaction'' \cite{perc2013evolutionary}. In this game, while only cooperators contribute to the common good, both cooperators and defectors benefit from it, that is, defectors are free-riders, social parasites. 

In the classical formulation of the PGG, for a constant individual cooperation cost, and a linear (as a function of the fraction of cooperators) benefit function, defection is the rational choice and constitutes the only Nash Equilibrium of the game. However, interior (mixed strategy) Nash equilibria can be found for nonlinear convex benefit functions \cite{motro1991co}. Other solutions proposed for the resilience of cooperation in the evolutionary dynamics of PGG include structured populations \cite{santos2008social,szolnoki2010reward} or information exchange \cite{gracia2014intergroup}.

In this paper, we investigate in detail how the strategies (cooperation or free-riding) of the PGG spread over the population of players under perhaps the simplest possible set of assumptions on the structure of agent contacts: Agents are partitioned in groups, each group constituting a fully-connected subpopulation within which its members exploit a common good, i.e., receive a payoff (benefit), that is a simple function of the abundance of cooperators in the group; this determines the individual fitness (reproductive power) of each agent. In order to implement the Darwinian competition for strategy spreading we use the ``well mixed'' myopic replicator dynamics (rule) \cite{gintis2000game}, i.e.: ``An agent imitates another, randomly chosen from the whole population, with a probability proportional to the fitness difference, if positive''. The model is presented in section \ref{section.model}. Two remarks are worth to make at this point: {\it a)} From the perspective of the multi-level selection debate, here we analyze a simple situation where the replicator entity is the individual, but on the other hand, the fitness of the replicator is ``group specific''; in other words the mesoscale (group level) is an essential component of the model itself. 
{\it b)} A math-analysis property of the ``replicator rule'' is its {\em threshold character} (``only imitate a fitter agent''). This ``non-analyticity'' allows simple arguments in the analysis of the model evolutionary dynamics.

In section \ref{section.results} we study the mesoscale description leading to a $m$-dimensional Markov process describing the evolution of the fraction of cooperators in each of the $m$ competing groups. The complete analysis of this dynamical system is greatly simplified by the use of symmetry arguments, provided the competing groups are equally sized (a generalization to unequal sized groups is studied in Appendix \ref{HeterogeneousSizeDistribution}). First, in \ref{2gr} and \ref{3gr}, we consider the cases of two and three competing groups where the exact analysis and visualization of the phase space portrait is greatly simplified by permutation symmetry considerations. These symmetry arguments turn out to be valid for any dimensions $m$, and we use them in \ref{mgr}. Our analysis reveals the existence (at the mesoscale level of description) of metastable symmetry-breaking macroscopic states. Exact bounds for ``meta-stability'' are moreover analytically found, due to the simplicity that remark {\em{b)}} above introduces in the analysis.

In section \ref{Higher} the agents' stochastic simulation for the evolution of the fraction of cooperators in each group shows a long term behavior in which the group cooperation values fluctuate around some well defined values, accurately predicted by the Markov dynamics analyzed in the previous section. Finally, in section \ref{section.conclusions}  we discuss the implications of the model in terms of the social norm, group selection, and sustainability of the common good, together with the conclusions.

\section{The model}
\label{section.model}

\begin{center}
\begin{figure}[h!]
    \centering
    \includegraphics[width=0.9\columnwidth]{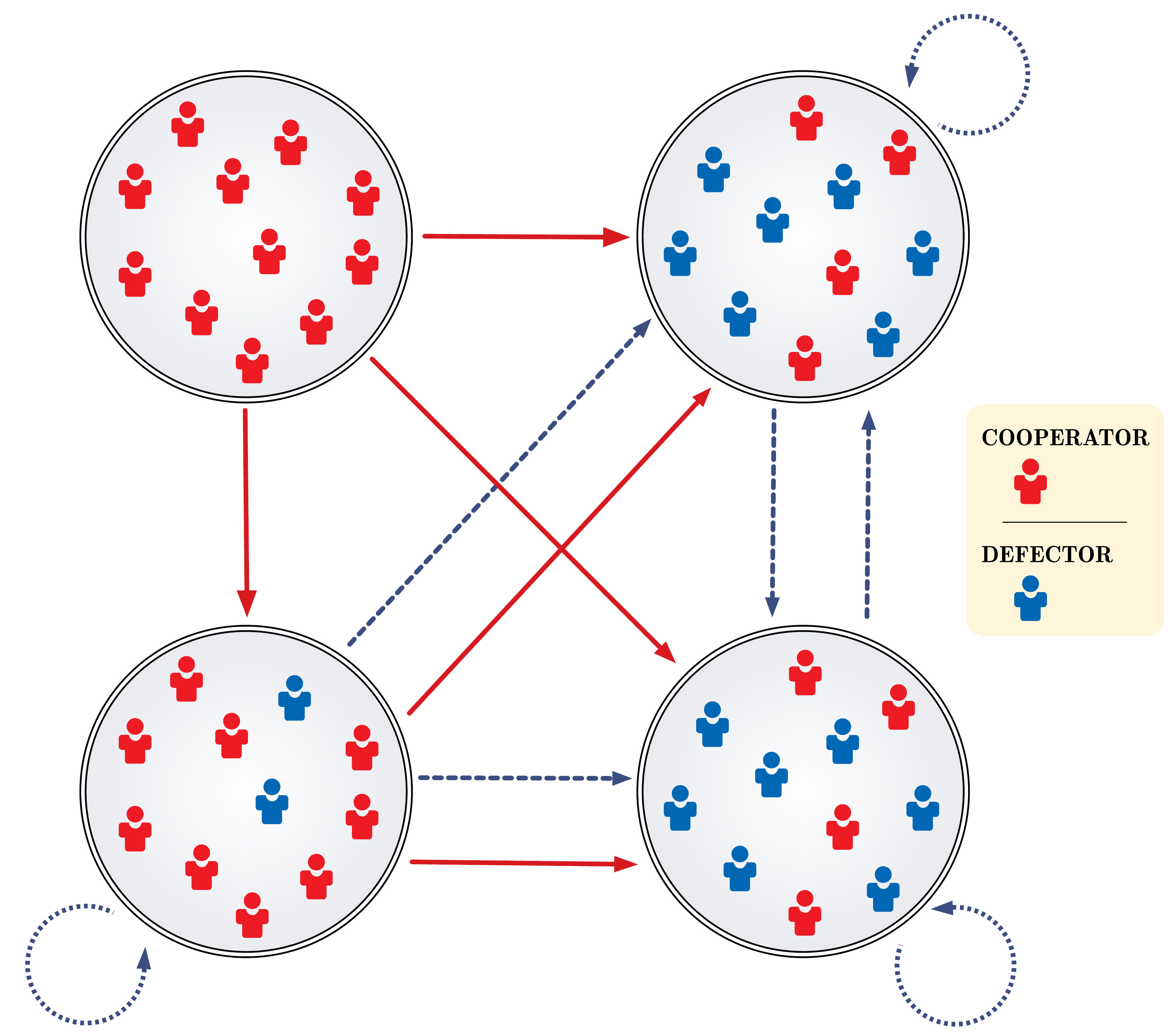}
    \caption{(Color online)  \textbf{Structure of the model.} Strategist agents are disposed in groups and can either cooperate or defect. Agents' payoffs are proportional to the number of cooperators in their group, cooperators having to pay an extra cost. Although agents obtain their payoff (fitness) from their group, imitation takes place from any group.
    In the diagram, solid red arrows represent the cooperation invasion flows, and blue dashed arrows the defection flows. See the text for further details.}
    \label{fig:model}
\end{figure}
\end{center}

This is a simple model of evolutionary game dynamics, where a population of strategists play a Public Goods game (PGG), from where they earn their reproductive power, i.e., their fitness. A population of $N=m \times n$ agents is divided into $m$ groups, each one with $n$ agents (Figure \ref{fig:model}). Inside each group, the agents play a Public Goods Game with two possible strategies; cooperate ($C$) and defect ($D$). Let us denote by $f_g^C$ and $f_g^D$ the payoff earned, respectively, by a cooperator and a defector in the group $g$ ($=1,\ldots,m$). If $n_g$ is the number of cooperators in group $g$ and the individual contribution to the public good is fixed to 1,
these payoffs are, in the standard linear PGG:

\begin{equation}
f_g^C=\frac{rc_g}{n}-1\;\;,\;\;\;\;\;\; f_g^D=\frac{rc_g}{n} \; \;,
\label{ind_payoff}
\end{equation}
\noindent where the slope $r>1$ of the payoff is often called ``synergy factor''. The fitness of an individual is 
assumed to be proportional to his payoff. Equations (\ref{ind_payoff}) can be rewritten as:
\begin{equation}
f_g^C=rp_g-1\;\;,\;\;\;\;\;\; f_g^D=rp_g \; \;,
\label{ind_payoff_p}
\end{equation}
where $p_g$ is the fraction of cooperator agents in group $g$.

To implement the Darwinian (natural selection) competition for strategic reproduction/spread, we choose the discrete version of the myopic replicator dynamics \cite{gintis2000game,santos2005scale}. On one hand, its continuum limit for pairwise games is, straightaway \cite{gintis2000game}, the well-studied replicator equation \cite{hofbauer2003evolutionary}. On the other, its threshold character, see equation (\ref{replicatorDynamics}) below, i.e. ``never change state by imitation of a less fitted agent'', turns out to be a source of simplicity in the analysis, as we will see below.

The idea of this model es pretty clear, namely, fitness comes from the group, while imitation takes place anywhere. Specifically, at each time step, which represents one generation of the discrete evolutionary time, all the agents play a one-shot PGG and obtain a payoff. After that, the individuals synchronously update their strategies in the following way: each agent $i$ compares its payoff with that of a random agent $j$ chosen equiprobably from any group, including own. Subsequently, if agent $j$ has a lower payoff than agent $i$, this keep his/her strategy, while if it is higher $i$ imitates $j$'s strategy with a probability $\Pi^{j\rightarrow i}$ proportional (wherefrom the term ``myopic'') to the payoff difference:

\begin{equation}
\Pi^{j\rightarrow i}=\frac{f^j-f^i}{\Delta f_{max}} \,\theta(f^j-f^i)\;\;,
\label{replicatorDynamics}
\end{equation}

\noindent where $\theta$ is the Heaviside step function ($\theta(y)=1$ for $y>0$ and $\theta(y)=0$ for $y\leq 0$), and $\Delta f_{max}$ is an arbitrary strict upper bound for the possible difference of agent payoff. It fixes the characteristic time scale. 

Note that for the simplest partition, where the game is played by all agents altogether, i.e. $m=1$, one realizes easily from equation (\ref{ind_payoff}) that the extinction of the cooperative strategy is the only possible evolutionary outcome. Free-riding is an unbeatable strategy.

\section{Markov phase space analysis.}
\label{section.results}

In this section we will consider equal sized groups of $n$ individuals. As there is no networked structure of agents interactions (neither regarding imitation rule, nor inside groups regarding payoff earning), the fractions (rational values, i.e. $l/n$ ($l=0,1,\cdots, n$)) $p_g$ ($g=1,\cdots, m$) specify the ``relevant'' (regarding dynamics) instantaneous description of the system's state of cooperation. However, please note that, if the term micro-state refers to the specification of the particular state of each agent, and these are distinguishable individuals, each particular value of $p_g=l/n$ represents a different number of micro-states, say $C_l^n$.

The phase space is a ($1/n$ lattice constant) grid over the $m$-dimensional unit hypercube. The stochastic population dynamics introduced in section \ref{section.model} defines a Markov process in this phase space, the model dynamics. 

We show below the analysis of the model dynamics for the simplest cases, say $m=2$ (\ref{2gr}) and $3$ (\ref{3gr}), for which a detailed geometrical investigation is feasible to visualize. The arguments used in the analysis of these explicitly solvable cases, are however easily seen to be valid for general values of the number $m$ of groups involved. We take advantage of the simplicity that permutation symmetry (interchange of group labels) considerations introduce in the analysis of the general $m$ case in \ref{mgr}. 

In section \ref{Higher} stochastic simulations results are interpreted to the light of the previous phase space non-linear analysis.

\subsection{Two groups}
\label{2gr}

\begin{center}
\begin{figure}[h!]
    \centering
    \includegraphics[width=0.9\columnwidth]{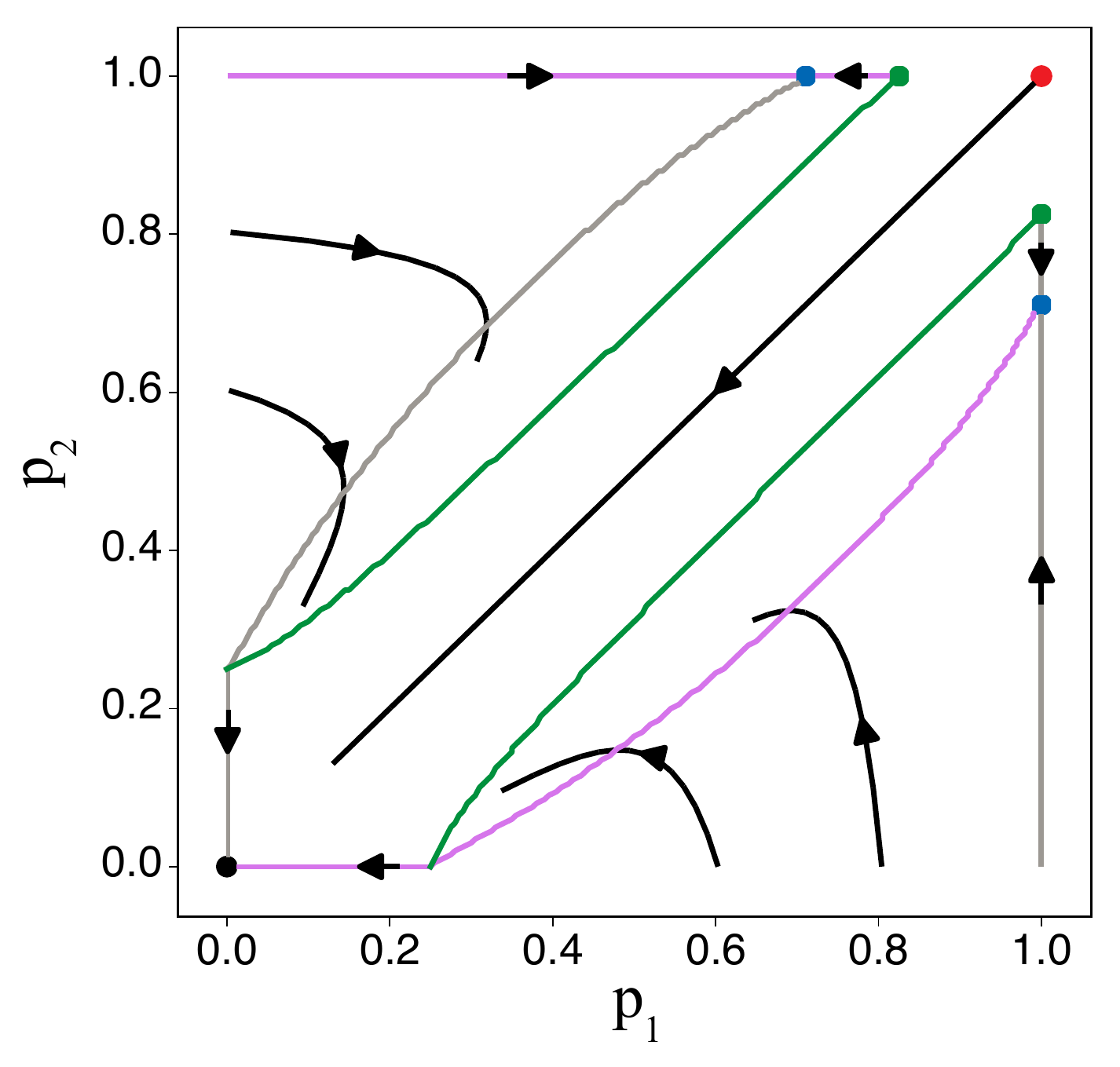}
    \caption{(Color online)  \textbf{Phase portrait for the two groups case.} A red dot indicates an unstable fixed point, a black dot a stable one, and a blue dot indicates a saddle point. The green dots correspond to
    the points $(1,p_2^{th})$ and $(p_1^{th},2)$. Gray (\textit{resp.}, purple) lines correspond to nullclines $dp_1=0$ ($dp_2=0$), while arrows represent the trajectories. Here, $r=4$. See the text for further details.}
    \label{fig:phase}
\end{figure}
\end{center}

Here, our phase space is the unit square. For equal sized groups, the invariance by interchange of group label (meaning that nothing at all changes if labels $1$ and $2$ are interchanged everywhere) allows us to restrict attention to the simplex (triangle) $0 \leq p_2\leq p_1 \leq 1$. It is simple to realise that the flow points outwards nowhere on the triangle boundary. Notwithstanding this invariance, we will show in the illustrating figure \ref{fig:phase} a ``full'' phase space portrait where this symmetry, at a first glance, can be easily acknowledged. 

The stochastic dynamics defined in previous section above for the agents' state evolution gives the following probabilities for the four possible (changes) relevant outcomes of the time step, say, increase or decrease of the number of cooperators in either group. They are functions of the fractions $p_i$ ($i=1,2$) of cooperators.

\begin{flushleft}
\begin{eqnarray}
P_1^+ &=& 0 \; , \nonumber \\
P_1^- &=& \displaystyle\frac{p_1}{\tau}\big((1-p_1) + (1-p_2)(f_2^D-f_1^C)\; \,\theta(f_2^D-f_1^C)\big) \; , \nonumber\\
P_2^+ &=& \displaystyle\frac{1-p_2}{\tau}\;p_1(f_1^C-f_2^D) \; \,\theta(f_1^C-f_2^D) \; , \nonumber \\
P_2^- &=& \displaystyle\frac{p_2}{\tau}\left((1-p_1)(f_1^D-f_2^C) + (1-p_2)\right) \; ,\nonumber \\
\label{eq.PPPP}
\end{eqnarray}
\end{flushleft}
where $\theta$ is the Heaviside step function ($\theta(y)=1$ for $y>0$ and $\theta(y)=0$ for $y\leq 0$), $\tau$ is the characteristic time scale (here, $\tau = 2 \Delta f_{max}$), and {\em e.g.}, $P^+_1$ is the probability of the transition $p_1 \rightarrow p_1+1/n$, while $P^-_2$ stands for the probability of decreasing $p_2$ (by $1/n$), etc ... In the previous formulas has been applied $f_i^D-f_i^C=1$. The flow on this ($p_1, p_2$) unit square  is ($i = 1, 2$):

\begin{equation}
dp_i = \frac{1}{n}(P_i^+ - P_i^-) \;.\label{dp_i}
\end{equation}

First we locate the nullclines, $dp_i=0$, on the triangle. 

\begin{itemize}

\item[\bf{1.}] {$dp_1=0$}.

This locus includes the corners $(0,0)$ and $(1,1)$, and the segment of the edge $p_1=1$ of $p_2$ values for which $f_1^C>f_2^D$. The conditions $f_1^C=f_2^D$ and $p_1=1$ determine the upper bound, $p_2^{th}$, of this branch of nullcline:
\begin{equation}
    p_{2} ^{th}= \frac{\sqrt{1+4r(r-1)} -1}{2r}\,.
\end{equation}
Note that for $p_2>p_{2} ^{th}$, $dp_1<0$, and the flow on the edge points inwards, while for $0 \leq p_2 <p_2^{th}$, the Heaviside function vanishes and the flow is co-linear to edge (a cooperator in group 1 does not imitate defectors in group 2).

\item[\bf{2.}] {$dp_2=0$}.

This includes the corners $(0,0)$ and $(1,1)$, and two branches. The first branch is the segment $0 \leq p_1 \leq 1/r$ on the edge $p_2=0$. The second branch is interior to the simplex, and its graph connects the points $(1/r, 0)$ and $(1, p_2^*)$, where $p^*_2$ is explicitly computed as
\begin{equation}
    p_2^* = \frac{r-1}{r+1} \;.
    \label{eq:fixpt2gps}
\end{equation}
 We see that this nullcline shows a singularity at $p_1=1/r$.

\end{itemize}

Due to the Heaviside $\theta$ functions in equation (\ref{eq.PPPP}), there is a line of singularities of the flow field, namely the intersection of the locus $f_1^C=f_2^D$ with the simplex. This is a curve connecting the points $(1/r, 0)$ and $(1, p_2^{th})$, where the constant-$dp_2$ isoclines show a singular behavior, similar to that of the $dp_2 = 0$ nullcline, that we have seen above.

The stationary states (fixed points) of the phase space flow have to be in the intersection of the nullclines. Thus there are three fixed points $(0,0)$, $(1,1)$, and $(1, p^*_2)$. 

In order to determine the stability properties of these fixed points, one should first compute the flow's Jacobian matrices at them, whose spectral decompositions (eigenvalues and eigen-subspaces) inform us (concisely) on their stability against perturbations in the linear regime. .

The $(i,j)$ Jacobian matrix's element, expresses how a variation in $p_j$, in the linear approximation, modifies $dp_i$, i.e. it is $$ J_{ij} = \frac{\partial dp_i}{\partial p_j}\; ,$$
and the results of the spectral analysis of the Jacobian matrix at the three fixed points (see Appendix \ref{appendix.two.groups.case} for details) can be summarized as follows:

\begin{itemize}

\item[{$(0,0)$}] Both eigenvalues are negative, as any fluctuation is damped out. Thus, complete defection is a local attractor. In fact, it is the global attractor for any initial condition in the interior of the simplex.

\item[{$(1,1)$}] Both eigenvalues are positive, meaning that cooperation in both groups is a repeller fixed point. Any perturbation (the appearance of a defector in either group) is amplified.

\item[ {$(1, p^*_2)$}] This is always a saddle point. Its stable manifold is the branch of the $dp_1=0$ nullcline, the segment $0 \leq p_2 <p_2^{th}$ on the edge. The unstable linear manifold is tangent to the interior branch of the $dp_2=0$ nullcline.
In Figure  \ref{fig:phase}, the saddle ($p_1=1,\; p_2=p_2^*$) is represented by a blue dot, and the upper point of the $dp_1=0$ nullcline, ($p_1=1,\; p_2=p_2^{th}$) by a green dot.

\end{itemize}

Figure \ref{fig:phase} displays the phase portrait for $r= 2$, where previous results can be checked by simple inspection.

\subsection{Three groups}
\label{3gr}

While we analyze here the case $m=3$, we also keep an eye on general $m$ values, because several conclusions from this analysis are easily seen to remain valid for an arbitrary large number of groups in the system.

The set of micro-states is the unit cube. One easily realizes that the fully defective corner $(0,0,0)$ is a fixed point of the dynamics. Also, it is easily seen that any small increase from zero in the fraction of cooperators in one or more groups, induces a restoring flow. Full defection is an absorbing state, an attractor. On the other hand, the other fully symmetric corner, the full cooperation corner, $(1,1,1)$ is also a fixed point, but it is unstable against defective fluctuations in one or more groups, and then is a repeller. The main diagonal connecting both, {\em i.e.}, the set of fully symmetric states, is a flow trajectory of strictly decreasing value of cooperation in every group. Indeed, after a little reflection, all this is true {\em mutatis mutandi} for any value of $m \geq 2$. In any dimension $m$ of the phase space, the set of fully symmetric micro-states is invariant and moreover, any small fluctuation orthogonal to it induces a restoring flow.

The remaining six corners of the unit cube are not fixed points, for at least one group is full defective (and at least one is full cooperative), and then a flow of increasing fraction of cooperators in the full defective group is ensured. Also this argument applies independently of the value of $m$.

Now we look for eventual fixed points on the twelve edges of the unit cube. Due to the symmetry by labels interchange, they are grouped into three classes of equivalence, that correspond to invariant subsets under symmetry transformations:

\begin{itemize}

\item The three axes are an invariant set under symmetry transformations. To fix ideas, think of the $p_3$-axis, at an abscissa $0<p_3$. The components of the flow orthogonal to this axis are both positive, thus we conclude that the are no fixed points on the axes other than the attractor at origin. Note that this was already clear for the two groups case; it is easy to realize that it is true for any number $m$ of groups involved.

\item The six cube edges that are neither adjacent to the origin nor to the $(1,1,1)$ corner ({\em e.g.} the segment $(0,p_2,1)$) form the second equivalence class. In these micro-states (points in these axes) there is one fully defective group and the flow is non-null due to the zero payoff of its defectors. Let us note that this simple consideration leads also to the conclusion that the three faces of the unit cube adjacent to origin cannot have on them a fixed point other than the origin.

\item The three cube edges adjacent to the full cooperation corner, say the segments $(p_1,1,1)$, $(1,p_2,1)$ and $(1,1,p_3)$, form the third class of edges. To fix ideas, think of {\em e.g.} $(1,p_2,1)$. For values of $p_2$ large enough, defectors in group 2 have larger payoff than cooperators in the fully cooperative groups 1 and 3, and there, the flow points towards the interior of the unit cube. If we denote by $p^{th}$ the value of $p_2$ for which $f_2^{D}=f_1^{C}=f_3^C$, and $p_1=p_3=1$, {\em i.e.}
\begin{equation}
    p ^{th}= \frac{r+1}{2r}\left(\sqrt{1+\frac{8r(r-1)}{(r+1)^2}} -1 \right)\,,
\end{equation}
the segment $0\leq p_2 \leq p^{th}$ is an invariant set, in other words, there the flow is co-linear to edge. Clearly, $dp_2>0$ at $p_2=0$, while at $p_2=p^{th}$ (where $f_2^{D}=f_1^{C}=f_3^C$) intra-group imitation leads to $dp_2<0$. Thus there is a fixed point $0<\hat{p}<p^{th}$ inside the segment, where the nullcline surface $dp_2=0$ intersects the edge $(1,p_2,1)$.

\end{itemize}

To proceed in the search for fixed points located at the phase space boundary, we have finally to consider the three faces adjacent to the fully cooperative corner $(1,1,1)$, that form a class of equivalent faces. To allow for analytics, we explicitly consider one of these, say the face defined by $p_3=1$, and, due to the symmetry by interchange of labels 1 and 2, we focus attention onto the triangular simplex $p_2 \leq p_1$, as in the previous subsection \ref{2gr}. 

Note that we have already inferred the existence of a fixed point $(1,\hat{p},1)$ located at the edge. The replicator dynamics defines the flow (\ref{dp_i}) on it with the following transition probabilities (where $\tau$ is the characteristic time scale, here $\tau = 3 \Delta f_{max}$):

\begin{eqnarray}
P_3^+ &=& 0  \nonumber \\
P_3^- &=& \displaystyle\frac{1}{\tau} \left(\sum_{i=1,2}(1-p_i)(f_i^D-f_3^C)\,\theta(f_i^D-f_3^C)\right) \; , \nonumber \\
P_1^+ &=& \displaystyle\frac{1-p_1}{\tau}(f_3^C-f_1^D)\,\theta(f_3^C-f_1^D) \, , \nonumber \\
P_1^- &= &\displaystyle\frac{p_1}{\tau}\left((1-p_1) + (1-p_2) (f_2^D-f_1^C)\,\theta(f_2^D-f_1^C)\right) \;, \nonumber \\
P_2^+ &=& \displaystyle\frac{1-p_2}{\tau} \left((f_3^C-f_2^D) \,\theta(f_3^C-f_2^D)\right.\nonumber\\
& &  \left.+\, p_1 (f_1^C-f_2^D) \,\theta(f_1^C-f_2^D) \right) \, , \nonumber \\
P_2^- &=& \displaystyle\frac{p_2}{\tau}\left( (1-p_2) + (1-p_1) (f_1^D-f_2^C)\,\theta(f_1^D-f_2^C) \right) \; . \nonumber \\
\end{eqnarray}

First we determine the region defined by $P_3^-=0$ (equiv. $f_1^D <f_3^C$), where the flow remains on the face. The condition $f_1^D =f_3^C$ defines a line $\tilde{p}_1(p_2)$ which intersects the edge $(p_1, 0, 1)$ at $\tilde{p}_1(0)$, and the symmetry line $(p, p, 1)$ at $\tilde{p}_1(p_2=p_1)$. Note that 
the condition $f_2^D <f_3^C$ is also satisfied due to our restriction to the $p_2 \leq p_1$ triangle.
The exact analytical expression $\tilde{p}(q)$, for the borderline $\tilde{p}_1(p_2)$ is 
\begin{equation}
\tilde{p}(q) = \frac{1+rq}{2r}\left(\sqrt{1+\frac{4r(r-1)(1+q)}{(1+qr)^2}} -1 \right)\,,
\label{ptilde}
\end{equation}
from which the previous intersection ($\tilde{p}(0)$, $\tilde{p}(\tilde{p})$, and $p^{th}= \tilde{p}(1)$) points can be explicitly determined as functions of the model parameters. 

Only to the left of this line the flow remains on the plane $p_3=1$. We then see that the invariant segment on the vertical edge, including the fixed point $(1,\hat{p},1)$, is disconnected from this region. On the contrary, the segment of the symmetry line below $\tilde{p}_1(p_2=p_1)$ is an invariant set included in the region, where we now focus attention. At the lower bound $(0,0,1)$ of this segment the flow is positive, while at the upper one $(\tilde{p}_1,\tilde{p}_1,1)$, cooperators in groups 1 and 2 imitate defectors, thus the flow is negative. Thus, there is a fixed point $(p^*,p^*,1)$ inside this segment where the nullcline surface $dp_2=0$ intersects the symmetry line. Note that being this line invariant, the condition $dp_2=0$ entails that also $dp_1=0$. In fact, the intersection of the $dp_1=0$ nullcline with the face has two isolated points, namely $(1,1,1)$ and $(p^*,p^*,1)$, and the segment $(1,0\leq p_2 \leq p^{th},1)$, which is isolated from the region $P_3^-=0$.

\begin{center}
\begin{figure}[ht]
    \centering
    \includegraphics[width=0.9\columnwidth]{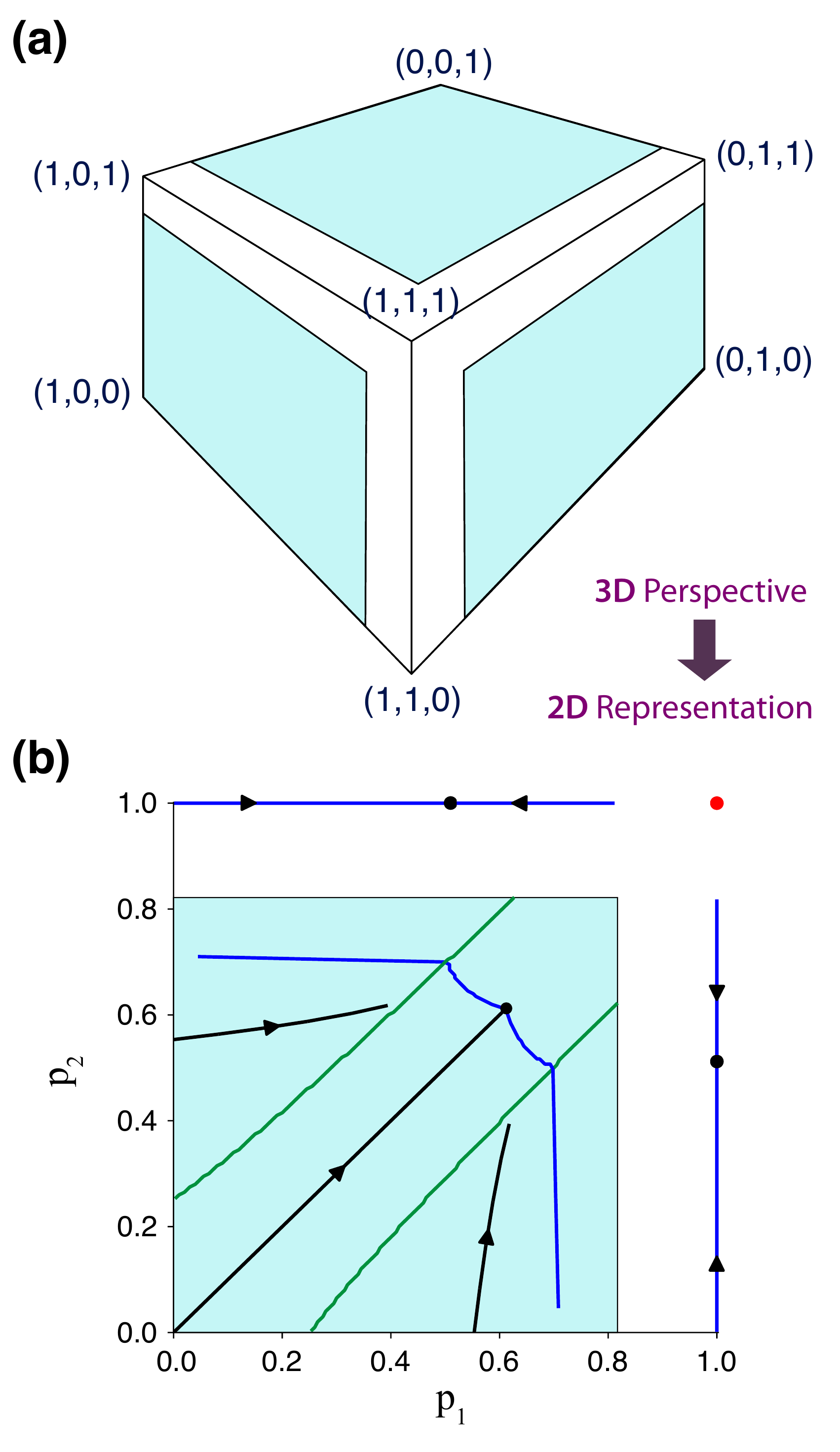}
    \caption{(Color online)  \textbf{Three groups case.}
    Top panel: Representation of the unit cube $(p_1, p_2, p_3)$. Colored areas correspond to the planar regions (co-dimension 1 stable manifolds) where one of the three groups is fully cooperator.  Bottom panel: Phase portrait for the three-groups case restricted to the plane $p_3=1$. The white area is the basin of the full defection state, while blue area corresponds to the stable manifold. The green lines correspond to the $f_1^C=f_2^D$ and $f_1^D=f_2^C$ singularities, blue lines to the nullclines, and arrows to trajectories. Inner black dot corresponds to the fixed
    point  $(p^*,p^*,1)$,
    and black dots located on the coordinate axes correspond to the fixed points $(\hat{p},1,1)$, $(1,\hat{p},1)$.
    In this plot, $r=4$. See the text for further details.}
    \label{fig:phase3D}
\end{figure}
\end{center}

We have found, regarding stationary states, that besides $(0,0,0)$ (attractor) and $(1,1,1)$ (repellor), there are six saddle fixed points: 
\begin{itemize}
\item Three of them are located on the three cube edges adjacent to $(1,1,1)$, at $\hat{p}$; each of them has a stable manifold on the segment $[0,p^{th}]$ over the corresponding edge; it is clear that fluctuations along directions orthogonal to the edge are repelled away the fixed point, so that its unstable manifold is two-dimensional. 
\item The last three fixed points are located at the symmetry lines of the three faces adjacent to $(1,1,1)$. The stable manifold for each of them is a two-dimensional (see top panel of Figure \ref{fig:phase3D} compact piece of the corresponding face. A fluctuation orthogonal to the face flows away, along the one-dimensional unstable manifold.
\end{itemize}

The three-dimensional visualization of the flow in the 3d phase space is dominated by the contraction of the interior phase space towards the fully defective state. However, located on three of the faces, there are co-dimension 1 invariant sets that are the stable manifolds of stationary states where one of the groups is fully cooperator, and the other two keep the same mixed state of strategic population. Also, located on the three cube edges, there are co-dimension 2 invariant segments that are the stable manifolds of stationary states with two full-C groups. We have arrived to these results through the use of exact and generalizable arguments.

\subsection{Is more (groups) different?}
\label{mgr}

The characterization of the deterministic trajectories in the phase space of our Markov model carried out for $m=2$ and $3$ in the previous subsections was obtained through arguments that are easily seen to hold for general values of $m \geq 2$, provided the $m$ groups are equally sized, and thus the symmetry by label interchange is preserved. Thus, the following educated conjecture can be safely put forward:
\begin{itemize}
\item[{\bf{ C1}}] For any value of $n_f$ ($1 \leq n_f \leq m-1$) there are $C^m_{n_f}$ saddle fixed points, where $n_f$ groups are full-C, and the rest ($m-n_f$) groups are mixed groups with a fraction of cooperators $p=\hat{p}(n_f, m)$. These states are equivalent under (label interchange) symmetry transformations. Any of them has a co-dimension $n_f$ ($m-n_f$ dimensional ) stable manifold.
\end{itemize}

Note that the value of $\hat{p}(n_f=1, m=2)$ was computed in \ref{2gr} as equation (\ref{eq:fixpt2gps}), while in \ref{3gr}, we denoted by $\hat{p}$ what in this general notation is $\hat{p}(n_f=2, m=3)$, and by $p^*$ what is now termed $\hat{p}(n_f=1, m=3)$.

The function $\hat{p}(n_f, m)$ can be easily obtained from the fixed point condition, as the positive solution (provided it is less than 1) to the quadratic equation 
\begin{equation}
    ap^2 + bp + c =0 \; ,
\label{eq:gdfixpt}
\end{equation}
with coefficients:
\begin{eqnarray}
    a &=& (m-n_f)(\frac{m}{n_f} +r -1) \;, \nonumber \\
    b &=& m -(r-1)(m-2n_f)  \;, \nonumber \\
    c &=&- (r-1)n_f \;. \nonumber
    \nonumber
\end{eqnarray}

One can quickly check that this result reproduces equation (\ref{eq:fixpt2gps}) for $m=2$ and $n_f=1$. No cooperator in the full $C$ groups will become a defector as long as its payoff is higher than the payoff of a defector in one of the ``mixed state'' groups. From this consideration, one easily finds a lower bound $p^{th}(m, n_f)$ for the ``fluctuation size'' {\em threshold of instability} of the full $C$ groups. The value of this lower bound, below which stability is ensured, is the positive root of the quadratic equation

\begin{equation}
    a'p^2 + b'p + c' =0 \; ,
\label{eq:thfixpt}
\end{equation}
with coefficients:
\begin{eqnarray}
    a' &=& (m-n_f)r  \;, \nonumber \\
    b' &=& (2n_f-m)r+(m-n_f)  \;, \nonumber \\
    c' &=&- (r-1)n_f \;. \nonumber
    \nonumber
\end{eqnarray}

This provides the exact functional dependence of our lower bound of instability threshold $p^{th}$ on all model parameters: $m$, $n_f$, and $r$. In the same way, equation (\ref{eq:gdfixpt}) provides the exact functional dependence of the mixed cooperation level $\hat{p}$ on these parameters. 

The mechanism that keeps a rather high average level of the groups is the positive contribution to $P^+_j$ (for mixed groups $j$) from the imitation of full $C$ group members. Simply said, the level of full cooperation ($p=1$) must be non-empty. Helas, this mechanism is fragile, for there is an instability threshold for fluctuations of the mixed groups cooperation, quantified by $p^{th}$, which restricts the stable manifolds of the fixed points to compact subsets on ($m - n_f$)-dimensional hypercubes . We will pay due attention to this fragility in the next subsection \ref{Higher}.

In the states that the conjecture {\bf{ C1}} refers to, there are groups with two different values for the fraction of cooperators, and one wonders if more than two values for $p_g$ are allowed in a fixed point.
We now provide an argument supporting that ``{\em more} (groups) {\em is different}'' \cite{anderson1972more}, regarding the fixed points of the Markov dynamics. More precisely, we will argue below that 
\begin{itemize}
\item[{\bf{ C2}}] There are metastable fixed points where groups with more than two different values of group cooperation $p_g$ coexist.

\end{itemize}

Consider one of the (type {\bf{ C1}}) stationary states with $n_f$ groups at the level $p=1$ and ($m-n_f$) groups at the ground level $p=\hat{p}$. Now, choose one (the focal group, now on) of these latter groups, and change its fraction of cooperators to a value $p^{up} > \hat{p}$, such that:
\begin{itemize}
\item $p^{up} < p^{th}$. This condition ensures that this change has no influence on the $n_f$ groups in the full $c$ level.
\item $p^{up}$ is high enough to make impossible the imitation by cooperators in the focal group of defectors in ground groups. 
\item The number $m$ of groups is large enough to render very small the effects of the focal group on the equilibrium value of the ground level.
\end{itemize}

\begin{center}
\begin{figure}[h!]
    \centering
    \includegraphics[width=0.9\columnwidth]{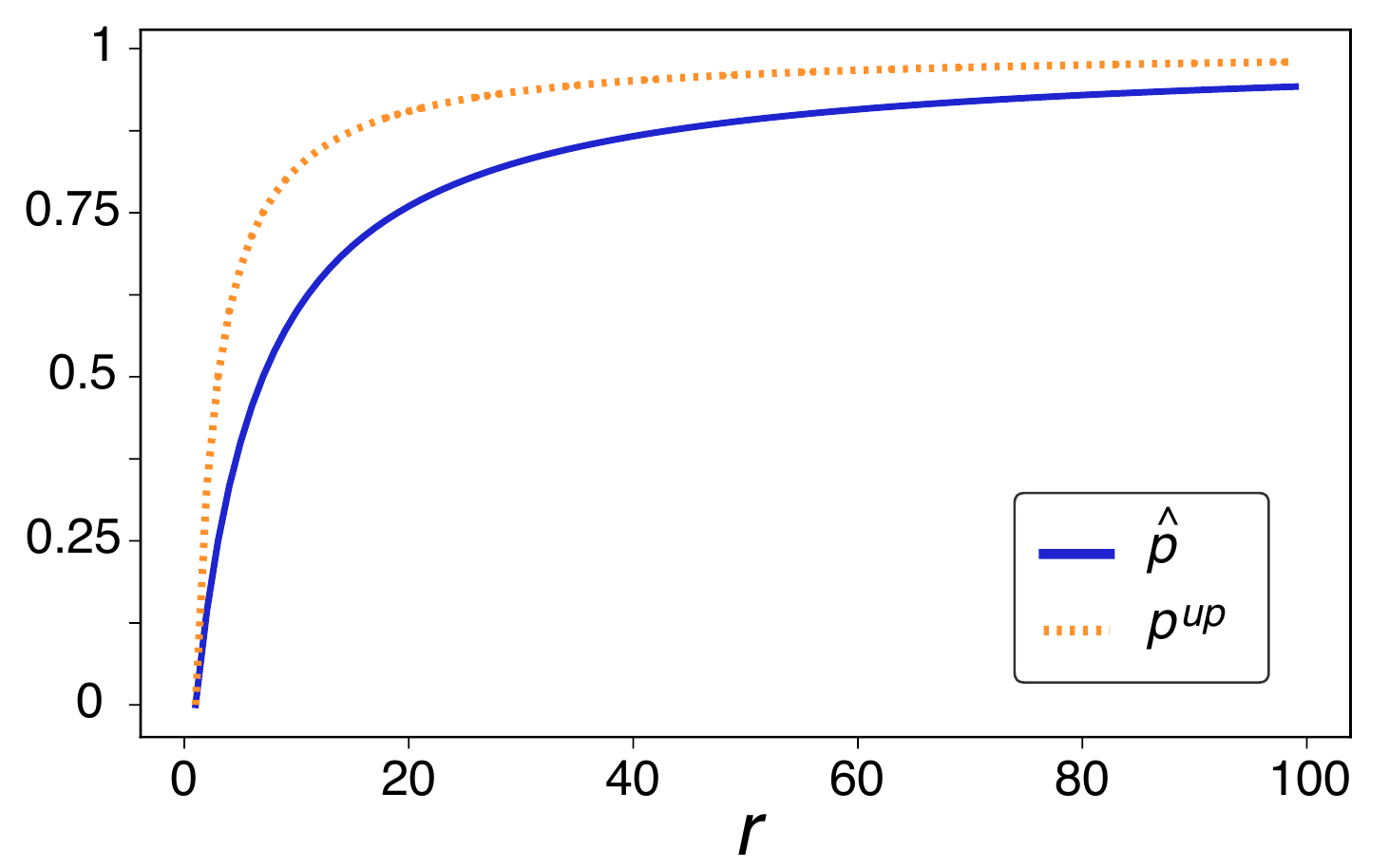}
    \caption{(Color online) \textbf{Ground and upper fixed points.} The graphs show the theoretical values of $\hat{p}$ (solid blue line) and $p^{up}$ (dashed orange line) as a function
    of $r$. The number of groups has been fixed to $m=6$, and the number of full-C groups to
    $n_f=1$. See the text for further details.}
\label{fig:FixedPoints}
\end{figure}
\end{center}

From the last two assumptions, one determines (see Appendix \ref{appendix_p^up}) an analytical expression for $p^{up}$ as a function of $\hat{p}$ and model parameters. Provided this value satisfies the first item above, this state can be taken as initial condition for the numerical direct integration of the Markov (either discrete or continuum) dynamics as a check of our third assumption, that hopefully refine both, $p^{up}$ and $\hat{p}$, values. The results of the
numerical direct integration are shown in Figure \ref{fig:FixedPoints}.

The observations (that are quite generic, regarding variation of parameters) fully support the existence of fixed points with three levels of group cooperation, as well as its metastable character (see section \ref{Higher}).

For a large number $m$ of groups, there is no apparent reason that can forbid the existence of metastable fixed points with $\nu>3$ levels of cooperation, at least for some range of model parameters and occupancies of the $\nu$ levels, provided the full cooperation level is not empty. Whenever two given levels are known, the condition that cooperators in the upper level have a higher payoff than free-riders in the lower level groups fixes a threshold value, below which a new intermediate level of cooperation can be (depending on ranges of parameters) proved for meta-stability.

We see how, in this model case, ``{\em more}'' (groups) gives new kinds of metastable fixed points, by further breaking the permutation symmetry, and then ``{\em is different}'' \cite{anderson1972more}. We should at this point emphasize that all the fixed points we have found along this section \ref{section.results}, other than ``all groups are full $C$ or full $D$'', are only invariant under a proper subset of group permutations. In other words, all of them are ``symmetry-breaking'' states. Also, all of them have at least one group of full cooperation.

\section{Agent-based stochastic simulations: Finite group size fluctuations and parametric sensitivity.}
\label{Higher}

In the previous sections we have analyzed the dynamics of the Markov model. This is different from, although intended to mimic, the stochastic dynamics of agents that was a part of the model formal definition in \ref{section.model}. Indeed, the Markov dynamics is the mesoscopic (group level) description. The $m$ fractions $p_g$ are collective (group) variables, while any micro-state of the system of agents is really specified by $m \times n$ binary values ($C$ or $D$). The number $\mu$ of agent micro-states that are associated to a point $\{p_g\}$ ($g=1, \cdots, m$) in the Markov phase space is 

$$\mu(\{p_g\}) = \Pi_{g=1}^m \left( \begin{array}{c} n \\ n p_g \end{array} \right)$$

\noindent  defining a non-uniform measure on the $m$-dimensional Markov phase space, which is highly concentrated at intermediate values of the group cooperation. When a stochastic evolution from an initial agent micro-state is visualized as a trajectory on the mesoscopic phase space, large amounts of information are lost.

Also, while the Markov dynamics is deterministic, the updating of agents strategy is stochastic. Boundaries in Markov phase space that keep deterministic trajectories inside invariant regions are easily crossed by stochastic trajectories  

\begin{center}
\begin{figure}[h!]
    \centering
    \includegraphics[width=0.9\columnwidth]{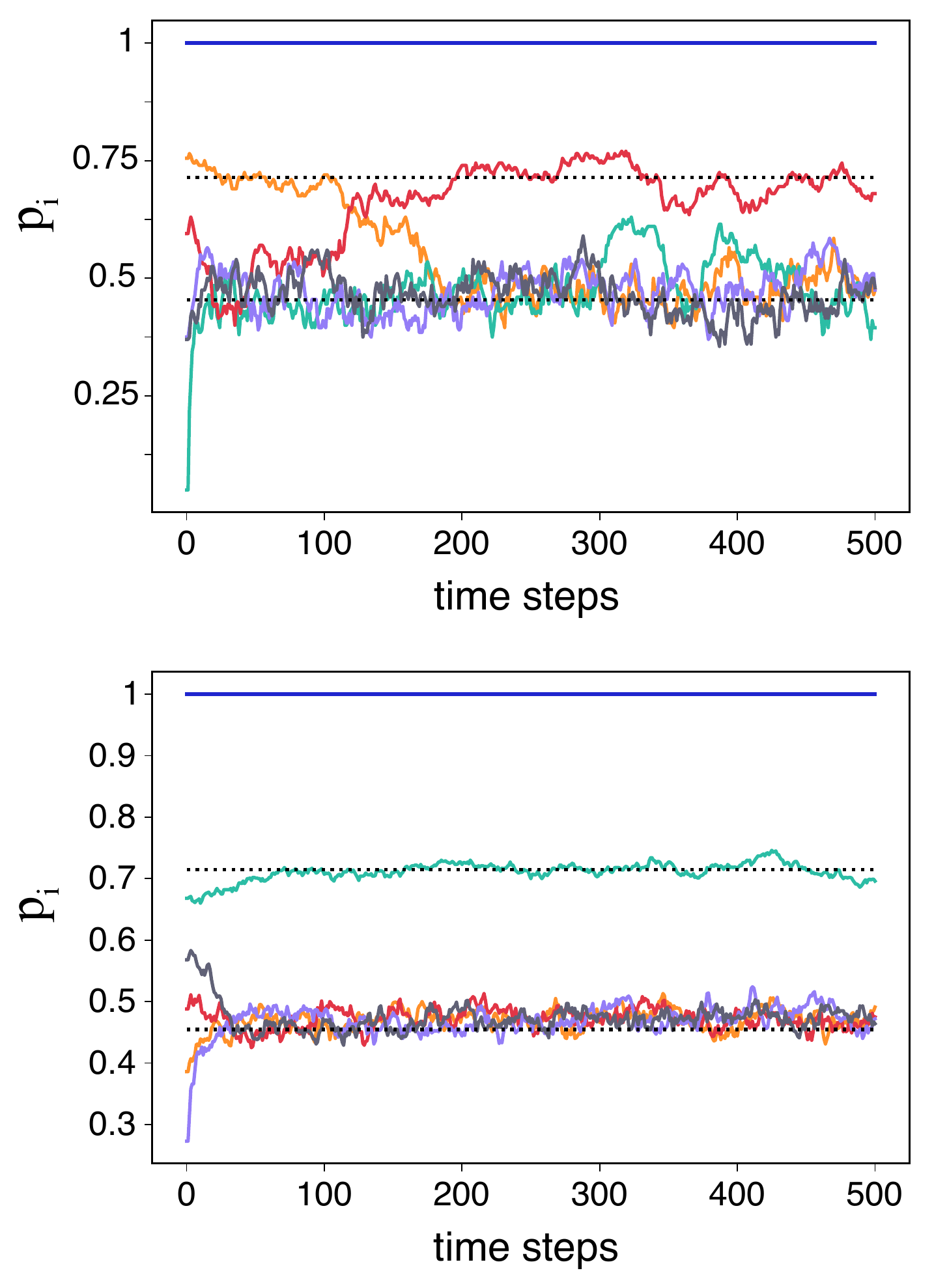}
    \caption{(Color online) \textbf{Stochastic evolution of the system.} The graphs show the time evolution of the fractions of cooperators for two representative realizations of the six-groups case, each solid line corresponding to a group. Dotted lines correspond to the theoretical predictions
    for $\hat{p}$ and $p^{up}$. For small group sizes (top panel, $N=200$) fluctuations allow groups to exchange their levels of cooperation, while for larger group sizes (bottom panel, $N=1000$) one group clearly detaches from the rest to occupy the upper fixed point, and the fluctuations are not large enough to allow exchange. In these plots, $r=6$.}
\label{fig:Evolution}
\end{figure}
\end{center}

\begin{center}
\begin{figure}[h!]
    \centering
    \includegraphics[width=0.9\columnwidth]{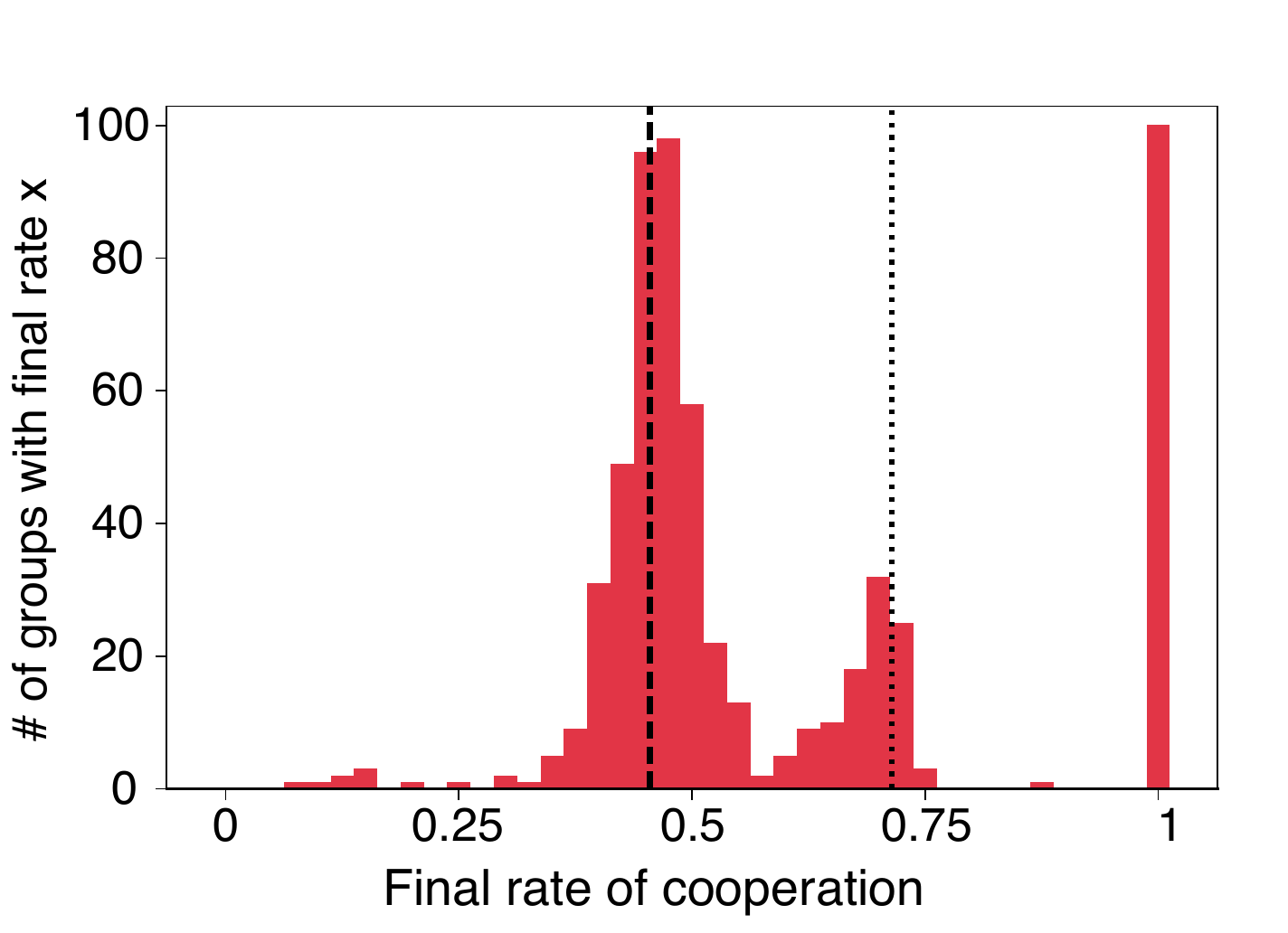}
    \caption{(Color online) \textbf{Histogram of cooperation level in groups.} The histogram counts the number of groups having a certain cooperation rate after 200 time steps. The counts have been accumulated over 100 independent realizations.  Dashed (\textit{resp.}, doted) line corresponds to the theoretical  predictions for $\hat{p}$ ($p^{up}$).  Here, the parameters are assigned the same values as in the upper panel of Figure \ref{fig:Evolution}: $m=6$, $r=6$, and $N=200$. The initial cooperation rates in the non full-C groups are given by a Gaussian determined by $\mu=\sigma^{2}=n/4$
}
\label{fig:Histogram}
\end{figure}
\end{center}

A convenient representation for trajectories of the many groups system is simply provided by the $m$ graphs for $p_g(t)$ ($g=1,\cdots, m$). In Figure
\ref{fig:Evolution} we show the time evolution of the
fraction of cooperators into the groups for two representative realizations
corresponding to different group sizes, $n=200$ (top) and $n=1000$ (bottom),
together with the theoretical values of $\hat{p}$ and $p^{up}$. As shown, the level of cooperation in the groups oscillates around points $\hat{p}$ and $p^{up}$. This fact is clearly displayed in Figure \ref{fig:Histogram}, which represents the histogram of the fraction of cooperative agents into the different groups, after the transient period and accumulated over 100 independent realizations, for the same values than those used in the upper panel of Figure \ref{fig:Evolution}.

The synergy factor $r$ allows the agents to get a higher payoff for the same contribution, and helps enhancing the cooperators over the defectors. By increasing $r$, a given strategy yields a higher payoff for the same cooperation frequency, eventually allowing a cooperator in a group with cooperative population $n_C + x$ to beat a defector in a group with cooperative population $n_C$; it substantially helps the cooperation to maintain itself in a many-groups setup. In general, a full-C group can resist invasion by defectors if $r > n$. In that case, a single defector in a group of size $n$, which would earn the highest payoff a defector could expect, would still receive a payoff lower than cooperator in a full-C cooperator. For $r>n$, a group reaching full cooperation will not leave this state, as long as the fluctuations are not large enough to allow the system to escape from that state.

It is interesting to look at the fluctuations to infer whether cooperation is sustainable or not in this setup. As expected and shown in Figure \ref{fig:Evolution}, fluctuations decay as the population size increases.
For high enough values of $n$, the system is stable; the fluctuations around
the fixed point do not allow a defector to beat a cooperator in the fully cooperative group, and cooperation is maintained. This remained true even for values of $r$ of the order of $n$: indeed, the higher the synergy factor, the less likely it becomes for a full-C cooperator to turn into a defector. We argue, by contrast, that finite-size effects can be dramatic: in some cases (typically $n \approx 10$), the fluctuations around the fixed point might allow a group to reach and overcome the critical population level, allowing invasion of the full-C group by defectors; the system can not be considered stable anymore in this case. Actually, the fluctuations follow the inverse square root of the population size (experimentally $1/2\sqrt{n}$), which gives us an estimate of the robustness of the system as a function of the size of the groups.

\section{Discussion and summary}
\label{section.conclusions}

In this study, we have taken a public goods approach to understand cooperative
behavior in group-structured populations. In the proposed model, individuals are
located in groups. The fitness of the individuals is related to their group performance, whereas the imitation takes place globally, that is, agents have a global vision and can imitate the most successful behaviors. We have shown that cooperation is maintained by a homogeneous group of cooperators. Note that this fully cooperative group does not necessarily have to be related to a physical group, but a social norm. In this sense, the existence of a cooperative social norm \cite{weber2008suckers} can be interpreted as a non-null probability of cooperation, which in the proposed model is mathematically equivalent to a homogeneous group of cooperators. This social norm can be based either on moral principles or driven by both empirical and normative expectations \cite{guth1994competition,bicchieri2005grammar,chudek2011culture}. 

The model is analytically solved through a Markovian approach, showing the existence of inner equilibria, invariant manifolds and thresholds for metastability. It is worth noting that although both individual and group levels are based on the well-mixed approach, the model exhibits intermediate rates of cooperation under the replicator dynamics. Furthermore, agent-based stochastic simulations show 
group cooperation values fluctuating around
the fixed points predicted by the Markov
dynamics, numerically validating the analytical predictions. Finally, this study has implications on the effect of social norms and group selection
on the sustainability of the commons.

\begin{acknowledgments}
We acknowledge partial support from Project UZ-I-2015/022/PIP, the Government of Arag\'on and FEDER funds, Spain through grant E36-17R to FENOL, and from MINECO and FEDER funds (grant FIS2017-87519-P). Y.M. also acknowledges partial support from Intesa Sanpaolo Innovation Center. The funders had no role in study design, data collection, and analysis, decision to publish, or preparation of the manuscript. 
\end{acknowledgments}

\section{Appendices}

\subsection{Appendix: Jacobian for the two groups case}
\label{appendix.two.groups.case}

In this appendix we present the general expression of the four terms of the Jacobian and the computation for $J_{22}$ evaluated around the fixed point. Let us set $F_1=f_2^D-f_1^C$ and $F_2=f_1^D-f_2^C$.
From (\ref{eq.PPPP}) and (\ref{dp_i}), it follows:

\begin{eqnarray}
J_{11}&=& p_1 -(1-p_2)\left(F_1 + p_1\left(\frac{\partial(F_1)}{\partial p_1}\right)\right)\,\theta(F_1)\;, \nonumber\\ \nonumber\\
J_{12}&=&p_1\left((1-p_2)\frac{\partial(F_1)}{\partial p_2}-F_1\right)\,\theta(F_1)\;, \nonumber\\ \nonumber\\
J_{21}&=&(1-p_2)\left(F_1 +p_1\frac{\partial(F_1)}{\partial p_1}\right)\,\theta(-F_1) \nonumber\\
&&-F_2 +(1-p_1)\frac{\partial(F_2)}{\partial p_1} \;, \nonumber\\
J_{22}&=&p_1\left((1-p_2)\frac{\partial F_1}{\partial p_2}-F_1\right)\,\theta(-F_1) -(1-p_1)F_2 \nonumber\\
&&-(1-p_2) -p_2\left((1-p_1)\frac{\partial F_2}{\partial p_2}\right)\;. \nonumber\\\nonumber
\end{eqnarray}

$J_{22}$ value can be evaluated around the fixed point $p^*$ computed in (\ref{eq:fixpt2gps}):

\begin{equation} \label{eq:J22@fixpt}
\begin{split}
J_{22}  =  2r\;\left(\frac{r+1}{r-1}-1\right)\nonumber\\
\end{split}
\end{equation}
which is always negative for $r \geq 0$.

\subsection{Appendix: estimation of the upper fixed point}
\label{appendix_p^up}

In this appendix we estimate the value of the metastable fixed point $p^{up}$ discussed in Point \textbf{C2} of Section \ref{mgr}. 
Let us consider more than two groups $m>2$
and, at least, one full-C group and more than
one mixed groups $1 \leq n_f \leq m-1$). The polynomial equation for the population in
the upper fixed point $p^{up}$ is found by 
imposing \textbf{C2} conditions
in the replicator dynamics, which drastically reduces the $P^-$ term and thus results in a higher value for the cooperation frequency:

\begin{equation}
    e(p^{up})^2 + gp^{up} + h =0 \; ,
\label{eq:upperfixpt}
\end{equation}
\begin{center}
    with:
\end{center}
\begin{eqnarray}
    e&=&n^{up}\left(\displaystyle\frac{n^{up}}{n_f} +r\right)  \;, \nonumber \\
    g&=&\alpha(n_f+\hat{n}\hat{p})-(r-1)n^{up} \;, \nonumber \\
    h&=&(1-r)(s+\hat{n}\hat{p}) \;, \nonumber
\end{eqnarray}
where $\hat{n}$ and $\hat{p}$ represent the number of groups in the lowest fixed point and their cooperation frequency, respectively, and $n^{up}$ and $p^{up}$ represent the number of groups in the upper fixed point and their cooperation frequency. After a first estimation of
those values based on the assumption that the variation of the ground value $\hat{p}$ is small, a better characterization of the fixed points can be obtained by refining the values of the fixed points and of the mean cooperation iteratively until convergence. The final value of the mean cooperation in mixed groups $\bar p$  is given by:

\begin{equation}
    \bar p=\displaystyle\frac{n^{up}p^{up} + (m-n_f-n^{up})\hat{p}}{m-n_f} \; .
    \label{eq:averagepredict}
\end{equation}

Since it is possible to compute the value of the mean cooperation frequency in the general case, we can compute the upper limit $p^{th}$ allowed for sustainability of cooperation. Let $f_C^f$ be the fitness of a cooperator in the full-C group, and $f_D^{th}$ be the fitness of a defector in the $p^{th}$ group.
From the limit condition:
\begin{equation}
f_C^f = f_D^{th} \\,
\end{equation}
it follows:
\begin{eqnarray} 
p^{th}r &=& (r-1)\;,\nonumber\\
p^{th} &=& 1-\frac{1}{r}\;.\nonumber\\
\label{eq:limitcase}
\end{eqnarray}

If in a group, the cooperation frequency overcomes this value, cooperators in the fully cooperative group can turn into defectors and the system would be in the basin of attraction of the fully defective state. 

\subsection{Appendix: Unequal sized groups}
\label{HeterogeneousSizeDistribution}

An interesting way towards the generalization of this model is to introduce disparity in groups sizes. Note that, considering that all the agents have the same probability to be chosen for imitation, larger groups will be more influential than smaller ones. This asymmetry may allow defectors in large groups to invade small cooperative groups.

As in the case of equal group sizes, without a fully cooperative group, the system is in the basin of attraction of the full-defection state. 

Since the imitation probabilities depend on the size of each group, the probabilities for a group $i$ to increase and decrease its fraction of cooperators $p_i$ by $1/n_i$ become, respectively:

\begin{eqnarray}
P_i^+ &=& \frac{n_i(1-p_i)}{\tau}\sum_{\substack{j=1 \\ j\neq i}}^m  n_jp_j(f_j^C-f_i^D)\,\theta(f_j^C-f_i^D)\;,	\nonumber\\
P_i^- &=& \frac{n_ip_i}{\tau}\sum_{j=1}^m n_jp_j(f_j^D-f_i^C)\,\theta(f_j^D-f_i^C)\;,	\nonumber \\
\end{eqnarray}
where $n_i$ represents the total number of agents of group $i$, $N$ represents the total number of agents in the whole system,
and $\tau$ is the characteristic time scale
involving $\Delta f_{max}$ and $N$.

The expression for the fixed point $p^*$
corresponding to all the mixed groups (non full-C groups) sharing the same value $p_i$ is obtained by setting $P_i^+ = P_i^-$:

\begin{eqnarray}
(N - n_f)^2p^*(1-p^*)=n_f(N - n_f)(1-p^*)(r-p^*r)\;,\nonumber\\
\end{eqnarray}
which yields the value of the fixed point:

\begin{equation}
p^* = \frac{n_f(r-1)}{N - n_f}\;.
\label{FixedPointHeterogeneous}
\end{equation}

A full-C group will resist invasion by defectors if:

\begin{equation} 
f_C^f > f_D^g \; \;
 \Rightarrow \; \; \frac{r-1}{r} > p_g^*n_g\;,
\label{eq:cdtionHeterogeneous}
 \end{equation}
where $n_g$ represents the size of the biggest group (excluding full-C group). 

Condition (\ref{eq:cdtionHeterogeneous}) relates the size of the fully cooperative group, the size of the biggest mixed group and the synergy factor $r$. It expresses whereas the full-C group can survive, and therefore, whether cooperation is stable in such a system or not.


\begin{thebibliography}{32}
\expandafter\ifx\csname natexlab\endcsname\relax\def\natexlab#1{#1}\fi
\expandafter\ifx\csname bibnamefont\endcsname\relax
  \def\bibnamefont#1{#1}\fi
\expandafter\ifx\csname bibfnamefont\endcsname\relax
  \def\bibfnamefont#1{#1}\fi
\expandafter\ifx\csname citenamefont\endcsname\relax
  \def\citenamefont#1{#1}\fi
\expandafter\ifx\csname url\endcsname\relax
  \def\url#1{\texttt{#1}}\fi
\expandafter\ifx\csname urlprefix\endcsname\relax\def\urlprefix{URL }\fi
\providecommand{\bibinfo}[2]{#2}
\providecommand{\eprint}[2][]{\url{#2}}

\bibitem[{\citenamefont{Smith et~al.}(1982)}]{smith1982evolution}
\bibinfo{author}{\bibfnamefont{J.~M.} \bibnamefont{Smith}}
  \bibnamefont{et~al.}, \emph{\bibinfo{title}{Evolution and the theory of
  games}} (\bibinfo{year}{1982}).

\bibitem[{\citenamefont{Pennisi}(2005)}]{pennisi2005did}
\bibinfo{author}{\bibfnamefont{E.}~\bibnamefont{Pennisi}},
  \bibinfo{journal}{Science} \textbf{\bibinfo{volume}{309}},
  \bibinfo{pages}{93} (\bibinfo{year}{2005}).

\bibitem[{\citenamefont{Chuang et~al.}(2009)\citenamefont{Chuang, Rivoire, and
  Leibler}}]{chuang2009simpson}
\bibinfo{author}{\bibfnamefont{J.~S.} \bibnamefont{Chuang}},
  \bibinfo{author}{\bibfnamefont{O.}~\bibnamefont{Rivoire}}, \bibnamefont{and}
  \bibinfo{author}{\bibfnamefont{S.}~\bibnamefont{Leibler}},
  \bibinfo{journal}{Science} \textbf{\bibinfo{volume}{323}},
  \bibinfo{pages}{272} (\bibinfo{year}{2009}).

\bibitem[{\citenamefont{Gore et~al.}(2009)\citenamefont{Gore, Youk, and
  Van~Oudenaarden}}]{gore2009snowdrift}
\bibinfo{author}{\bibfnamefont{J.}~\bibnamefont{Gore}},
  \bibinfo{author}{\bibfnamefont{H.}~\bibnamefont{Youk}}, \bibnamefont{and}
  \bibinfo{author}{\bibfnamefont{A.}~\bibnamefont{Van~Oudenaarden}},
  \bibinfo{journal}{Nature} \textbf{\bibinfo{volume}{459}},
  \bibinfo{pages}{253} (\bibinfo{year}{2009}).

\bibitem[{\citenamefont{Kagel and Roth}(2016)}]{kagel2016handbook}
\bibinfo{author}{\bibfnamefont{J.~H.} \bibnamefont{Kagel}} \bibnamefont{and}
  \bibinfo{author}{\bibfnamefont{A.~E.} \bibnamefont{Roth}},
  \emph{\bibinfo{title}{The handbook of experimental economics, volume 2: the
  handbook of experimental economics}} (\bibinfo{publisher}{Princeton
  university press}, \bibinfo{year}{2016}).

\bibitem[{\citenamefont{Kollock}(1998)}]{kollock1998social}
\bibinfo{author}{\bibfnamefont{P.}~\bibnamefont{Kollock}},
  \bibinfo{journal}{Annual review of sociology} \textbf{\bibinfo{volume}{24}},
  \bibinfo{pages}{183} (\bibinfo{year}{1998}).

\bibitem[{\citenamefont{Trivers}(1971)}]{trivers1971evolution}
\bibinfo{author}{\bibfnamefont{R.~L.} \bibnamefont{Trivers}},
  \bibinfo{journal}{The Quarterly review of biology}
  \textbf{\bibinfo{volume}{46}}, \bibinfo{pages}{35} (\bibinfo{year}{1971}).

\bibitem[{\citenamefont{Fowler}(2005)}]{fowler2005human}
\bibinfo{author}{\bibfnamefont{J.~H.} \bibnamefont{Fowler}},
  \bibinfo{journal}{Nature} \textbf{\bibinfo{volume}{437}}, \bibinfo{pages}{E8}
  (\bibinfo{year}{2005}).

\bibitem[{\citenamefont{Nowak and Sigmund}(1998)}]{nowak1998evolution}
\bibinfo{author}{\bibfnamefont{M.~A.} \bibnamefont{Nowak}} \bibnamefont{and}
  \bibinfo{author}{\bibfnamefont{K.}~\bibnamefont{Sigmund}},
  \bibinfo{journal}{Nature} \textbf{\bibinfo{volume}{393}},
  \bibinfo{pages}{573} (\bibinfo{year}{1998}).

\bibitem[{\citenamefont{Nowak and May}(1992)}]{nowak1992evolutionary}
\bibinfo{author}{\bibfnamefont{M.~A.} \bibnamefont{Nowak}} \bibnamefont{and}
  \bibinfo{author}{\bibfnamefont{R.~M.} \bibnamefont{May}},
  \bibinfo{journal}{Nature} \textbf{\bibinfo{volume}{359}},
  \bibinfo{pages}{826} (\bibinfo{year}{1992}).

\bibitem[{\citenamefont{Hamilton}(1964)}]{hamilton1964genetical}
\bibinfo{author}{\bibfnamefont{W.~D.} \bibnamefont{Hamilton}},
  \bibinfo{journal}{Journal of theoretical biology}
  \textbf{\bibinfo{volume}{7}}, \bibinfo{pages}{17} (\bibinfo{year}{1964}).

\bibitem[{\citenamefont{Wynne-Edwards}(1962)}]{wynne1962animal}
\bibinfo{author}{\bibfnamefont{V.~C.} \bibnamefont{Wynne-Edwards}},
  \bibinfo{type}{Tech. Rep.} (\bibinfo{year}{1962}).

\bibitem[{\citenamefont{Wilson}(1975)}]{wilson1975theory}
\bibinfo{author}{\bibfnamefont{D.~S.} \bibnamefont{Wilson}},
  \bibinfo{journal}{Proceedings of the national academy of sciences}
  \textbf{\bibinfo{volume}{72}}, \bibinfo{pages}{143} (\bibinfo{year}{1975}).

\bibitem[{\citenamefont{Wilson and Sober}(1994)}]{wilson1994reintroducing}
\bibinfo{author}{\bibfnamefont{D.~S.} \bibnamefont{Wilson}} \bibnamefont{and}
  \bibinfo{author}{\bibfnamefont{E.}~\bibnamefont{Sober}},
  \bibinfo{journal}{Behavioral and brain sciences}
  \textbf{\bibinfo{volume}{17}}, \bibinfo{pages}{585} (\bibinfo{year}{1994}).

\bibitem[{\citenamefont{Rapoport et~al.}(1965)\citenamefont{Rapoport, Chammah,
  and Orwant}}]{rapoport1965prisoner}
\bibinfo{author}{\bibfnamefont{A.}~\bibnamefont{Rapoport}},
  \bibinfo{author}{\bibfnamefont{A.~M.} \bibnamefont{Chammah}},
  \bibnamefont{and} \bibinfo{author}{\bibfnamefont{C.~J.}
  \bibnamefont{Orwant}}, \emph{\bibinfo{title}{Prisoner's dilemma: A study in
  conflict and cooperation}}, vol. \bibinfo{volume}{165}
  (\bibinfo{publisher}{University of Michigan press}, \bibinfo{year}{1965}).

\bibitem[{\citenamefont{Doebeli and Hauert}(2005)}]{doebeli2005models}
\bibinfo{author}{\bibfnamefont{M.}~\bibnamefont{Doebeli}} \bibnamefont{and}
  \bibinfo{author}{\bibfnamefont{C.}~\bibnamefont{Hauert}},
  \bibinfo{journal}{Ecology letters} \textbf{\bibinfo{volume}{8}},
  \bibinfo{pages}{748} (\bibinfo{year}{2005}).

\bibitem[{\citenamefont{Andreoni and Miller}(1993)}]{andreoni1993rational}
\bibinfo{author}{\bibfnamefont{J.}~\bibnamefont{Andreoni}} \bibnamefont{and}
  \bibinfo{author}{\bibfnamefont{J.~H.} \bibnamefont{Miller}},
  \bibinfo{journal}{The economic journal} \textbf{\bibinfo{volume}{103}},
  \bibinfo{pages}{570} (\bibinfo{year}{1993}).

\bibitem[{\citenamefont{Cooper et~al.}(1996)\citenamefont{Cooper, DeJong,
  Forsythe, and Ross}}]{cooper1996cooperation}
\bibinfo{author}{\bibfnamefont{R.}~\bibnamefont{Cooper}},
  \bibinfo{author}{\bibfnamefont{D.~V.} \bibnamefont{DeJong}},
  \bibinfo{author}{\bibfnamefont{R.}~\bibnamefont{Forsythe}}, \bibnamefont{and}
  \bibinfo{author}{\bibfnamefont{T.~W.} \bibnamefont{Ross}},
  \bibinfo{journal}{Games and Economic Behavior} \textbf{\bibinfo{volume}{12}},
  \bibinfo{pages}{187} (\bibinfo{year}{1996}).

\bibitem[{\citenamefont{Gracia-L{\'a}zaro
  et~al.}(2012)\citenamefont{Gracia-L{\'a}zaro, Ferrer, Ruiz, Taranc{\'o}n,
  Cuesta, S{\'a}nchez, and Moreno}}]{gracia2012heterogeneous}
\bibinfo{author}{\bibfnamefont{C.}~\bibnamefont{Gracia-L{\'a}zaro}},
  \bibinfo{author}{\bibfnamefont{A.}~\bibnamefont{Ferrer}},
  \bibinfo{author}{\bibfnamefont{G.}~\bibnamefont{Ruiz}},
  \bibinfo{author}{\bibfnamefont{A.}~\bibnamefont{Taranc{\'o}n}},
  \bibinfo{author}{\bibfnamefont{J.~A.} \bibnamefont{Cuesta}},
  \bibinfo{author}{\bibfnamefont{A.}~\bibnamefont{S{\'a}nchez}},
  \bibnamefont{and} \bibinfo{author}{\bibfnamefont{Y.}~\bibnamefont{Moreno}},
  \bibinfo{journal}{Proceedings of the National Academy of Sciences}
  \textbf{\bibinfo{volume}{109}}, \bibinfo{pages}{12922}
  (\bibinfo{year}{2012}).

\bibitem[{\citenamefont{Perc et~al.}(2013)\citenamefont{Perc,
  G{\'o}mez-Garde{\~n}es, Szolnoki, Flor{\'\i}a, and
  Moreno}}]{perc2013evolutionary}
\bibinfo{author}{\bibfnamefont{M.}~\bibnamefont{Perc}},
  \bibinfo{author}{\bibfnamefont{J.}~\bibnamefont{G{\'o}mez-Garde{\~n}es}},
  \bibinfo{author}{\bibfnamefont{A.}~\bibnamefont{Szolnoki}},
  \bibinfo{author}{\bibfnamefont{L.~M.} \bibnamefont{Flor{\'\i}a}},
  \bibnamefont{and} \bibinfo{author}{\bibfnamefont{Y.}~\bibnamefont{Moreno}},
  \bibinfo{journal}{Journal of the royal society interface}
  \textbf{\bibinfo{volume}{10}}, \bibinfo{pages}{20120997}
  (\bibinfo{year}{2013}).

\bibitem[{\citenamefont{Motro}(1991)}]{motro1991co}
\bibinfo{author}{\bibfnamefont{U.}~\bibnamefont{Motro}},
  \bibinfo{journal}{Journal of Theoretical Biology}
  \textbf{\bibinfo{volume}{151}}, \bibinfo{pages}{145} (\bibinfo{year}{1991}).

\bibitem[{\citenamefont{Santos et~al.}(2008)\citenamefont{Santos, Santos, and
  Pacheco}}]{santos2008social}
\bibinfo{author}{\bibfnamefont{F.~C.} \bibnamefont{Santos}},
  \bibinfo{author}{\bibfnamefont{M.~D.} \bibnamefont{Santos}},
  \bibnamefont{and} \bibinfo{author}{\bibfnamefont{J.~M.}
  \bibnamefont{Pacheco}}, \bibinfo{journal}{Nature}
  \textbf{\bibinfo{volume}{454}}, \bibinfo{pages}{213} (\bibinfo{year}{2008}).

\bibitem[{\citenamefont{Szolnoki and Perc}(2010)}]{szolnoki2010reward}
\bibinfo{author}{\bibfnamefont{A.}~\bibnamefont{Szolnoki}} \bibnamefont{and}
  \bibinfo{author}{\bibfnamefont{M.}~\bibnamefont{Perc}}, \bibinfo{journal}{EPL
  (Europhysics Letters)} \textbf{\bibinfo{volume}{92}}, \bibinfo{pages}{38003}
  (\bibinfo{year}{2010}).

\bibitem[{\citenamefont{Gracia-Lazaro et~al.}(2014)\citenamefont{Gracia-Lazaro,
  Gomez-Gardenes, Floria, and Moreno}}]{gracia2014intergroup}
\bibinfo{author}{\bibfnamefont{C.}~\bibnamefont{Gracia-Lazaro}},
  \bibinfo{author}{\bibfnamefont{J.}~\bibnamefont{Gomez-Gardenes}},
  \bibinfo{author}{\bibfnamefont{L.~M.} \bibnamefont{Floria}},
  \bibnamefont{and} \bibinfo{author}{\bibfnamefont{Y.}~\bibnamefont{Moreno}},
  \bibinfo{journal}{Physical Review E} \textbf{\bibinfo{volume}{90}},
  \bibinfo{pages}{042808} (\bibinfo{year}{2014}).

\bibitem[{\citenamefont{Gintis}(2000)}]{gintis2000game}
\bibinfo{author}{\bibfnamefont{H.}~\bibnamefont{Gintis}},
  \emph{\bibinfo{title}{Game theory evolving: A problem-centered introduction
  to modeling strategic behavior}} (\bibinfo{publisher}{Princeton university
  press}, \bibinfo{year}{2000}).

\bibitem[{\citenamefont{Santos and Pacheco}(2005)}]{santos2005scale}
\bibinfo{author}{\bibfnamefont{F.~C.} \bibnamefont{Santos}} \bibnamefont{and}
  \bibinfo{author}{\bibfnamefont{J.~M.} \bibnamefont{Pacheco}},
  \bibinfo{journal}{Physical Review Letters} \textbf{\bibinfo{volume}{95}},
  \bibinfo{pages}{098104} (\bibinfo{year}{2005}).

\bibitem[{\citenamefont{Hofbauer and Sigmund}(2003)}]{hofbauer2003evolutionary}
\bibinfo{author}{\bibfnamefont{J.}~\bibnamefont{Hofbauer}} \bibnamefont{and}
  \bibinfo{author}{\bibfnamefont{K.}~\bibnamefont{Sigmund}},
  \bibinfo{journal}{Bulletin of the American Mathematical Society}
  \textbf{\bibinfo{volume}{40}}, \bibinfo{pages}{479} (\bibinfo{year}{2003}).

\bibitem[{\citenamefont{Anderson}(1972)}]{anderson1972more}
\bibinfo{author}{\bibfnamefont{P.~W.} \bibnamefont{Anderson}},
  \bibinfo{journal}{Science} \textbf{\bibinfo{volume}{177}},
  \bibinfo{pages}{393} (\bibinfo{year}{1972}).

\bibitem[{\citenamefont{Weber and Murnighan}(2008)}]{weber2008suckers}
\bibinfo{author}{\bibfnamefont{J.~M.} \bibnamefont{Weber}} \bibnamefont{and}
  \bibinfo{author}{\bibfnamefont{J.~K.} \bibnamefont{Murnighan}},
  \bibinfo{journal}{Journal of personality and social psychology}
  \textbf{\bibinfo{volume}{95}}, \bibinfo{pages}{1340} (\bibinfo{year}{2008}).

\bibitem[{\citenamefont{G{\"u}th and Kliemt}(1994)}]{guth1994competition}
\bibinfo{author}{\bibfnamefont{W.}~\bibnamefont{G{\"u}th}} \bibnamefont{and}
  \bibinfo{author}{\bibfnamefont{H.}~\bibnamefont{Kliemt}},
  \bibinfo{journal}{Metroeconomica} \textbf{\bibinfo{volume}{45}},
  \bibinfo{pages}{155} (\bibinfo{year}{1994}).

\bibitem[{\citenamefont{Bicchieri}(2005)}]{bicchieri2005grammar}
\bibinfo{author}{\bibfnamefont{C.}~\bibnamefont{Bicchieri}},
  \emph{\bibinfo{title}{The grammar of society: The nature and dynamics of
  social norms}} (\bibinfo{publisher}{Cambridge University Press},
  \bibinfo{year}{2005}).

\bibitem[{\citenamefont{Chudek and Henrich}(2011)}]{chudek2011culture}
\bibinfo{author}{\bibfnamefont{M.}~\bibnamefont{Chudek}} \bibnamefont{and}
  \bibinfo{author}{\bibfnamefont{J.}~\bibnamefont{Henrich}},
  \bibinfo{journal}{Trends in cognitive sciences}
  \textbf{\bibinfo{volume}{15}}, \bibinfo{pages}{218} (\bibinfo{year}{2011}).

\end{thebibliography}

\end{document}